\documentclass[prd]{revtex4}

\usepackage[dvips]{epsfig}

\begin{document}

\title{Vacuum energies due to delta-like currents: simulating classical objects along branes with arbitrary codimensions}

\author{F. A. Barone\footnote{e-mail: fbarone@unifei.edu.br} and
G. Flores-Hidalgo\footnote{e-mail: gfloreshidalgo@unifei.edu.br}}
\affiliation{ICE - Universidade Federal de Itajub\'a, Av. BPS 1303, Caixa Postal 50 - 37500-903, Itajub\'a, MG, Brazil.}

\date{}

%%%%%%%%%
\begin{abstract}
	In this paper we investigate the vacuum energies of several models of quantum fields interacting with static external currents (linear couplings) concentrated along parallel branes with an arbitrary number of codimensions. We show that we can simulate the presence of static charges distributions as well as the presence of classical static dipoles in any dimension for massive and massless fields. We also show that we can produce confining potentials with massless self interacting scalar fields as well as long range anisotropic potentials. 
\end{abstract}
%%%%%%%%%

\maketitle

\baselineskip=20pt

%%%%%%%%%
\section{Introduction}
%%%%%%%%%

	It is well known in the literature that the coupling of a bosonic field (spin $0$ or spin $1$) with two static external currents with the shape of Dirac's delta functions concentrated at specific points of space gives rise to the Coulomb potential for massless fields, and the Yukawa potential for massive fields \cite{Zee,Itzykson,R}. This fact is one of the great triumphs of the Quantum Field Theory and enables us to comprehend, at the quantum level, the interaction between classical objects by describing them with Dirac's delta functions.
	
	Some natural questions that arise from these results concern the coupling between quantum fields and external static currents concentrated at specific regions of space.
We can mention, for instance, what might be the physical meaning of a spinorial static current with the shape of a Dirac's delta function,
or which kind of systems can be described by the interaction between bosonic fields with other types of external currents spatially concentrated, instead of currents with the shape of a Dirac's delta function. Even for the simple, and well known, system composed by two point-like Dirac's delta functions in $3+1$ dimensions, which leads to the Yukawa potential, we can ask about the influence due to additional typical quantum interactions, like the $\lambda\phi^{4}$ model in the case of scalar field.

	In this paper, which is devoted to the study of some of these questions, we calculate the vacuum energies for several models of quantum fields in interaction with external static currents concentrated at distinct parallel branes with arbitrary number of codimensions. Specifically, we have treated the cases in which static delta like currents interact linearly with massive or massless fields. We would like to point out that systems of quantum fields interacting with external potentials concentrated along branes (quadratic couplings in the quantum fields) have been treated in the literature from time to time and in different contexts, see for instance \cite{bordag,milton,S} and references cited therein, but, as far as the authors know, the coupling of quantum fields with external currents concentrated along branes (linear couplings in the quantum fields) is not a well explored subject.
	
	The paper is structured as follows;	in section (\ref{1}) we calculate the vacuum energy of a real scalar field with mass in $d+D+1$ dimensions and coupled to $N$ external Dirac-like currents concentrated at distinct regions of space. That is a generalization of some results which can be found in the literature for $3+1$ dimensions \cite{Zee} and are important calculations also because we establish, in a simple system, notation and the mathematical tools used in the whole paper. In section (\ref{2}) we consider the scalar field interacting with an external current with the shape of Dirac's delta functions derivatives. The results obtained are interesting once they show that the model considered can be used to describe the interaction between electric static dipoles in $d+D+1$ dimensions. Section (\ref{3,4}) is devoted to extend the previous results to the electromagnetic field, that brings the previous results to more realistic contexts. In section (\ref{5}) we consider a model in $d+D+1$ dimensions described by the fermionic field interacting with a static external current with the shape of Dirac's delta functions. We show that, for the massless field in $3+1$ dimensions and a current composed by two point-like Dirac's delta functions, the vacuum energy exhibits a similar spatial behavior in comparing with the electric field produced by a static electric dipole. In section (\ref{lambdaphi4}) we investigate the corrections due to the $\lambda\phi^{4}$ self-interaction on the results obtained in sections (\ref{1}) and (\ref{2}) for the restricted case with $3+1$ dimensions and currents concentrated along two branes. We show that we can produce confining potentials in first order in $\lambda$ and, as a particular case, the correction to the coulombian potential is proportional to the distance between the charges. Section (\ref{conclusao}) is devoted for some final comments and conclusions. 
	
	Along the paper we shall deal with models in $d+D+1$ dimensions and use Minkowski coordinates with the diagonal metric $(1,-1,-1,...,-1)$.
	
	The time coordinate shall be represented by $x^{0}$ and the four-vector position shall be designated by
%%%%%%%%%
\begin{equation}
\label{def4vetor}
x=(x^{0},x^{1},...,x^{d},x^{d+1},...,x^{d+D})\ .
\end{equation}
%%%%%%%%%
We shall also use the following notations
%%%%%%%%%
\begin{eqnarray}
\label{defxperpx|}
{\bf x}_{\perp}&=&(x^{1},...,x^{d})\nonumber\\
{\bf x}_{\|}&=&(x^{d+1},...,x^{d+D})\ ,
\end{eqnarray}
%%%%%%%%%
and similar ones for the momentums $k$.

%%%%%%%%%
\section{Scalar Field in $d+D+1$ dimensions with delta-like currents: charges distributions}
\label{1}
%%%%%%%%%

	In this section we study the vacuum energy of a system composed by the real scalar field in $D+d+1$ dimensions with mass and coupled linearly with $N$ $d$-dimensional delta functions concentrated at different regions of space. The model is described by the Lagrangian density
%%%%%%%%%
\begin{equation}
\label{modelo1}
{\cal L}_{I}=\frac{1}{2}(\partial_{\mu}\phi)(\partial^{\mu}\phi)-\frac{1}{2}m^{2}\phi^{2}+\Biggl(\sum_{p=1}^{N}\sigma_{p}\ \delta^{d}({\bf x}_{\perp}-{\bf a}_{p})\Biggr)\phi\ ,
\end{equation}
%%%%%%%%%
where we have $N$ fixed $d$-dimensional spatial vectors ${\bf a}_{p}$, $p=1,...,N$ and ${\bf x}_{\perp}$ is defined in (\ref{defxperpx|}). The parameters $\sigma_{p}$ are the coupling constants between the field and the delta functions.

	Notice that the model (\ref{modelo1}) can be interpreted as the scalar field interacting with the external current
%%%%%%%%%
\begin{equation}
\label{defJ}
J_{I}(x)=\sum_{p=1}^{N}\sigma_{p}\ \delta^{d}({\bf x}_{\perp}-{\bf a}_{p})\ ,
\end{equation}
%%%%%%%%%
(or a sum of external currents) so the functional generator of the Green's functions can be written as
%%%%%%%%%
\begin{equation}
\label{rfv}
{\cal Z}_{I}=\exp\Biggl(-\frac{i}{2}\int\int\ d^{d+D+1}x\ d^{d+D+1}y\ \ J_{I}(x)\Delta_{F}(x,y)J_{I}(y)\Biggr)\ ,
\end{equation}
%%%%%%%%%
where $\Delta_{F}(x,y)$ is the Feynman propagator
%%%%%%%%%
\begin{equation}
\label{defDeltaF}
\Delta_{F}(x,y)=\int{\frac{d^{d+D+1}k}{(2\pi)^{d+D+1}}\ \frac{\exp\Bigl[ik(x-y)\Bigr]}{k^{2}-m^{2}}}\ .
\end{equation}
%%%%%%%%%

	The functional generator of any quantum system, at the limit $T\to\infty$, can be written in the form \cite{Zee,Itzykson} 
%%%%%%%%%
\begin{equation}
\label{FG}
{\cal Z}=\exp(-iET)\ ,
\end{equation}
%%%%%%%%%
where $E$ is the lowest energy of the system.

	Comparing Eq's (\ref{rfv}) and (\ref{FG}), this last one with $E=E_{I}$ and ${\cal Z}={\cal Z}_{I}$, we have
%%%%%%%%%
\begin{equation}
\label{rfv1}
E_{I}=\lim_{T\rightarrow 0}\frac{1}{2T}\int\int\ d^{d+D+1}x\ d^{d+D+1}y\ \ J_{I}(x)\Delta_{F}(x,y)J_{I}(y)\ .
\end{equation}
%%%%%%%%%

	Substituting (\ref{defDeltaF}) and (\ref{defJ}) into (\ref{rfv1}) leads to
%%%%%%%%%
\begin{equation}
\label{asd}
E_{I}=\sum_{p=1}^{N}\sum_{q=1}^{N}\sigma_{p}\sigma_{q}\frac{1}{2T}{\cal I}_{p,q}\ ,
\end{equation}
%%%%%%%%%
where ${\cal I}_{p,q}$ is the following integral
%%%%%%%%%
\begin{eqnarray}
\label{AdefcalI}
{\cal I}_{p,q}=\int\int d^{D+d+1}x\ d^{D+d+1}y\ \delta^{(d)}({\bf x}_{\perp}-{\bf a}_{p})\Delta_{F}(x,y)\delta^{(d)}({\bf y}_{\perp}-{\bf a}_{q})\ . 
\end{eqnarray}
%%%%%%%%%

	Expression (\ref{asd}) contains terms where $p=q$ which account for contributions for the energy due to the self interactions of each delta function with itself. These terms must be discarded once, each of them, can be interpreted as the self energy of a given brane and do not contribute to the force acting on any brane	\footnote{A similar situation occurs in the Casimir effect where the self energy of the plates must be discarded\cite{milton,bordag}.}. So the energy due, strictly, to the interaction between the deltas is given by
%%%%%%%%%
\begin{eqnarray}
\label{asd1}
E_{I}&\to&\sum_{p=1}^{N}\sum_{q=1}^{N}\sigma_{p}\sigma_{q}\frac{1}{2T}{\cal I}_{p,q}-\sum_{p=1}^{N}\sigma_{p}\sigma_{p}\frac{1}{2T}{\cal I}_{p,p}\cr\cr
&=&\sum_{p=1}^{N}\sum_{q=1}^{N}\sigma_{p}\sigma_{q}(1-\delta_{pq})\frac{1}{2T}{\cal I}_{p,q}\ ,
\end{eqnarray}
%%%%%%%%%
where $\delta_{pq}$ is the Kronecker delta.

	In the appendix A the integral defined in (\ref{AdefcalI}) is simplified to the form
%%%%%%%%%
\begin{equation}
\label{AresultadocalI}
{\cal I}_{p,q}=-T\ L^{D}\ \int\frac{d^{d}{\bf k}_{\perp}}{(2\pi)^{d}}\ \frac{1}{{\bf k}_{\perp}^{2}+m^{2}}\ \exp{(-i{\bf k}_{\perp}\cdot{\bf a}_{pq})}\ ,
\end{equation}
%%%%%%%%%
where we defined
%%%%%%%%%
\begin{equation}
{\bf a}_{pq}={\bf a}_{p}-{\bf a}_{q}\ .
\end{equation}
%%%%%%%%%

	Substituting the result (\ref{AresultadocalI}) in Eq. (\ref{asd1}) we can write
%%%%%%%%%
\begin{equation}
\label{asd2}
{\cal E}_{I}=\frac{E_{I}}{L^{D}}=-\frac{1}{2}\sum_{p=1}^{N}\sum_{q=1}^{N}\sigma_{p}\sigma_{q}(1-\delta_{pq})\int\frac{d^{d}{\bf k}_{\perp}}{(2\pi)^{d}}\ \frac{1}{{\bf k}_{\perp}^{2}+m^{2}}\ \exp{(-i{\bf k}_{\perp}\cdot{\bf a}_{pq})}\ ,
\end{equation}
%%%%%%%%%
where we denoted by ${\cal E}_{I}$ the vacuum energy per unit parallel volume $L^{D}=\int d^{D}{\bf y}_{\|}$.

	From now on it shall be convenient to study, separately, the cases with and without mass.	For the massive case the integral which appears in Eq. (\ref{asd2}) is ill defined for higher dimensions, so we shall search for its analytic extension in such a way to obtain a well defined expression for the energy (\ref{asd2}) valid for higher dimensions as well. For this task we use the results of appendix B (\ref{Id}) and (\ref{Bprincipal}), this last one with $\nu=(d/2)-1$ and ${\bf p}={\bf k}_{\perp}$, in order to write
%%%%%%%%%
\begin{eqnarray}
\label{Icommassa}
\int d^{d}{\bf k}_{\perp}\ \frac{1}{{\bf k}_{\perp}^2+m^2} \exp(\pm i{\bf k}_{\perp}\cdot{\bf a})&=&(2\pi)^{d/2}a^{2-d}\int_{0}^{\infty}du\frac{u^{d/2}J_{(d/2)-1}(u)}{u^2+(ma)^{2}}\nonumber\\
&=&(2\pi)^{d/2}m^{d-2}(ma)^{1-(d/2)}K_{(d/2)-1}(ma)\ ,\ 0<d<5\ ,\ m>0\ ,\nonumber\\
&=&(2\pi)^{d/2}m^{d-2}G_{d}(ma)\ ,\ 0<d<5\ ,\ m>0\ ,
\end{eqnarray}
%%%%%%%%%
where $K_{n}(x)$ designates the $K$-Bessel function and we defined
%%%%%%%%%
\begin{equation}
\label{defG}
G_{d}(x)=x^{1-(d/2)}\ K_{(d/2)-1}(x)\ .
\end{equation}
%%%%%%%%%

	Notice that, even though the result (\ref{Icommassa}) had been obtained with the restriction $0<d<5$, the right hand side of the last line of (\ref{Icommassa}) is well defined for any integer $d$, so it is an analytic extension, valid for any integer $d$, for the integral on the left hand side of the first line\footnote{In fact, we are interested in $d=1,2,3,...$ .} in such a way that
%%%%%%%%%
\begin{equation}
\label{ext1}
\int d^{d}{\bf k}_{\perp}\ \frac{1}{{\bf k}_{\perp}^2+m^2} \exp(\pm i{\bf k}_{\perp}\cdot{\bf a})\rightarrow(2\pi)^{d/2}m^{d-2}G_{d}(ma)\ ,\ m>0\ .
\end{equation}
%%%%%%%%%
	
	Defining the variables $a_{pq}=|{\bf a}_{pq}|=|{\bf a}_{p}-{\bf a}_{q}|$ and substituting the integral which appears in (\ref{asd2}) by its analytic extension (\ref{ext1}) the energy (\ref{asd2}) reads
%%%%%%%%%
\begin{equation}
\label{calEphicommassa1}
{\cal E}_{I}(\sigma_{p},\sigma_{q},a_{pq},m\not=0,d,N)=-\frac{1}{2}\frac{m^{d-2}}{(2\pi)^{d/2}}\sum_{p=1}^{N}\sum_{q=1}^{N}\sigma_{p}\sigma_{q}(1-\delta_{pq})G_{d}(ma_{pq})\ ,
\end{equation}
%%%%%%%%%
which is valid for any integer $d$.

	The force per unit parallel volume $L^{D}$ acting on a given  current localized, for instance, by the perpendicular vector ${\bf a}_{\ell}$ can be calculated from expression (\ref{calEphicommassa1}) as follows 
%%%%%%%%%
\begin{eqnarray}
\label{fphideltacommassa}
{\bf\cal F}_{I,(\ell)}&=&-\Biggl(\frac{\partial}{\partial a_{\ell k}}\sum_{{k=1}\atop{k\not=\ell}}^{N}{\cal E}_{I}(\sigma_{p},\sigma_{q},a_{pq},m,d,N)\Biggr)\ \frac{{\bf a}_{\ell k}}{a_{\ell k}}\cr\cr
&=&-\frac{1}{(2\pi)^{d/2}}\sum_{{k=1}\atop{k\not=\ell}}^{N}\ \frac{\sigma_{\ell}\sigma_{k}}{a_{\ell k}^{d-1}}\ (ma_{\ell k})^{d/2}K_{d/2}(ma_{\ell k})\ \frac{{\bf a}_{\ell k}}{a_{\ell k}}\ ,
\end{eqnarray}
%%%%%%%%%
where we used the fact that
%%%%%%%%%
\begin{equation}
\frac{\partial G_{d}(x)}{\partial x}=-x^{1-(d/2)}K_{d/2}(x)\ 
\end{equation}
%%%%%%%%% 
and discarded a term $(1-\delta_{k\ell})$ once the summation does not include the terms where $k=\ell$. 
It is interesting to notice that the force (\ref{fphideltacommassa}) satisfies the superposition principle for the delta-like currents.

The energy for the massless case could be obtained by taking $m=0$ in Eq (\ref{calEphicommassa1}), which is well defined for $d\not=2$, as shall be exposed later. For complentess we shall perform the calculations with $m=0$ following a different approach.
		
	For the massless case let us consider, separately, the situations where $d=2$ and $d\not=2$. When $d\not=2$ we use the results obtained in appendix B (\ref{Id}) and (\ref{Bsegunda}), this last one with $\mu=(d/2)-2$, $\nu=(d/2)-1$ and ${\bf p}={\bf k}_{\perp}$, in order to write the integral which appears in Eq. (\ref{asd2}), for $m=0$, in the form
%%%%%%%%%
\begin{eqnarray}
\label{Isemmassa}
\int d^{d}{\bf k}_{\perp}\ \frac{1}{{\bf k}_{\perp}^2} \exp(\pm i{\bf k}_{\perp}\cdot{\bf a})&=&(2\pi)^{d/2}a^{2-d}\int_{0}^{\infty}du\ u^{(d/2)-2}J_{(d/2)-1}(u)\nonumber\\
&=&(2\pi)^{d/2}\ 2^{(d/2)-2}\ \Gamma\Biggl(\frac{d}{2}-1\Biggr)\ a^{2-d}\ ,\ 2<d<5\ ,
\end{eqnarray}
%%%%%%%%%

	Similarly to what happened for (\ref{Icommassa}), the result (\ref{Isemmassa}) was obtained with the restriction $2<d<5$, but the result in the second line of (\ref{Isemmassa}) is valid for $d\not=2,0,-2,-4,...$, so it is the analytic extension for the integral on the left hand side of (\ref{Isemmassa}) and we can write
%%%%%%%%%
\begin{equation}
\label{ext2}
\int d^{d}{\bf k}_{\perp}\ \frac{1}{{\bf k}_{\perp}^2} \exp(\pm i{\bf k}_{\perp}\cdot{\bf a})\rightarrow
(2\pi)^{d/2}\ 2^{(d/2)-2}\ \Gamma\Biggl(\frac{d}{2}-1\Biggr)\ a^{2-d}\ ,\ d\not=2\ ,
\end{equation}
%%%%%%%%%
where we considered only $d\geq1$.
	
	Substituting the integral which appears in (\ref{asd2}) with $m=0$ by its analytic extension (\ref{ext2}) we have the energy
%%%%%%%%%
\begin{equation}
\label{asd3}
{\cal E}_{I}(\sigma_{p},\sigma_{q},a_{pq},m=0,d\not=2,N)=-\frac{\Gamma\Bigl((d/2)-1\Bigr)2^{(d/2)-3}}{(2\pi)^{d/2}}\sum_{p=1}^{N}\sum_{q=1}^{N}\frac{\sigma_{p}\sigma_{q}}{a_{pq}^{d-2}}\ (1-\delta_{pq})\  .
\end{equation}
%%%%%%%%%

	Before we study the case where $m=0$ and $d=2$, let us investigate the behavior of the analytic extension (\ref{ext1}) for any $d\geq1$ and $m=0$. For this task we have to use the fact that \cite{Arfken}
%%%%%%%%%
\begin{eqnarray}
\label{Km->0}
K_{\nu}(z)&\stackrel{z=0}{\longrightarrow}&\frac{\Gamma(\nu)2^{\nu-1}}{z^{\nu}}\ ,\ \nu\not=0\cr\cr
K_{0}(z)&\stackrel{z=0}{\longrightarrow}&-\ln\Biggl(\frac{z}{2}\Biggr)-\gamma\ \ ,
\end{eqnarray}
%%%%%%%%%	
where $\gamma$ is the Euler constant.

	Substituting the first Eq. (\ref{Km->0}) in (\ref{defG}) and (\ref{ext1}) we are taken to (\ref{ext2}). Similarly, when $m=0$ the energy (\ref{calEphicommassa1}) is reduced to Eq. (\ref{asd3}). So the result (\ref{calEphicommassa1}) can be considered valid also when $m=0$ and $d\not=2$.
	
	For the situation where $m=0$ and $d=2$ the energy (\ref{calEphicommassa1}) exhibits divergences which do not depend on the distance between the branes and can be removed with no relevance for dynamics of the system. For this task we substitute the second Eq. (\ref{Km->0}) in  and (\ref{calEphicommassa1}) what leads to
%%%%%%%%%
\begin{equation}
\label{asd3d=2a}
{\cal E}_{I}(\sigma_{p},\sigma_{q},a_{pq},m\to0,d=2,N)=\frac{1}{4\pi}\sum_{p=1}^{N}\sum_{q=1}^{N}\sigma_{p}\sigma_{q}(1-\delta_{pq})\ln\Biggl(\frac{a_{pq}}{a_{0}}\Biggr)
+\frac{1}{4\pi}\sum_{p=1}^{N}\sum_{q=1}^{N}\sigma_{p}\sigma_{q}(1-\delta_{pq})\Biggl[\ln\Biggl(\frac{ma_{0}}{2}\Biggr)+\gamma\Biggr]\ ,
\end{equation}
%%%%%%%%%
where $a_{0}$ is a non vanishing arbitrary constant with dimension of length.

	In the limit $m\to0$ the last summation on the right hand side of (\ref{asd3d=2a}) is composed by divergent $a_{pa}$-independent terms, which do not contribute to interacting forces, so it can be discarded whith no relevance for dynamics,
%%%%%%%%%
\begin{equation}
\label{asd3d=2}
{\cal E}_{I}(\sigma_{p},\sigma_{q},a_{pq},m=0,d=2,N)=\frac{1}{4\pi}\sum_{p=1}^{N}\sum_{q=1}^{N}\sigma_{p}\sigma_{q}(1-\delta_{pq})\ln\Biggl(\frac{a_{pq}}{a_{0}}\Biggr)\ .
\end{equation}
%%%%%%%%%

	The calculations for the massless case with $d=2$ can be interpreted as an insertion of a mass as a regulator parameter introduced in order to remove $a$-independent divergences. 
	
	The force in the massless cases is obtained following the same steps which lead to the result (\ref{fphideltacommassa}). So, from (\ref{asd3}) and (\ref{asd3d=2}) we obtain
%%%%%%%%%
\begin{equation}
\label{fphideltasemmassa}
{\bf\cal F}_{I,(\ell)}=-\frac{\Gamma(d/2)2^{(d/2)-1}}{(2\pi)^{d/2}}\sum_{{k=1}\atop{k\not=\ell}}^{N}\ \frac{\sigma_{\ell}\sigma_{k}}{a_{\ell k}^{d-1}}\ \frac{{\bf a}_{\ell k}}{a_{\ell k}}\ ,
\end{equation}
%%%%%%%%%
which is valid for any value of $d$.

	As a check of consistence we can take the limit of vanishing mass in Eq. (\ref{fphideltacommassa}) by using Eq's (\ref{Km->0}). The results are given by the force (\ref{fphideltasemmassa}) for any value of $d$.

	The situations with $3+1$ dimensions and $N=2$ are of special interest; when $d=3$, $D=0$ and $N=2$ we have two point-like currents in $3+1$ dimensions, whose interaction energy is the Yukawa potential which can be obtained from (\ref{calEphicommassa1}) \cite{Zee}
%%%%%%%%%
\begin{equation}
E_{I}(a,m,d=3,D=0,N=2)=-\frac{\sigma_{1}\sigma_{2}}{4\pi}\frac{\exp(-ma)}{a}\ ,
\end{equation}
%%%%%%%%%
where we suppressed the sub-indices $1,2$ for $a$, once there is only one distance involved. For $m=0$ the above result leads to the Coulombian interaction.

	For $d=2$, $D=1$ and $N=2$ we have two delta-like currents concentrated along two different parallel strings placed at a distance $a$ from each other. In this case the energy per string length reads
%%%%%%%%%
\begin{equation}
\label{asd4}
{\cal E}_{I}(a,m,d=2,D=1,N=2)=-\frac{\sigma_{1}\sigma_{2}}{2\pi}K_{0}(ma)\ ,
\end{equation}
%%%%%%%%%
which is reduced, in the case $m=0$, to the expression
%%%%%%%%%
\begin{equation}
\label{asd5}
{\cal E}_{I}(a,m=0,d=2,D=1,N=2)=\frac{\sigma_{1}\sigma_{2}}{2\pi}\ln\biggl(\frac{a}{a_{0}}\biggr)\ ,
\end{equation}
%%%%%%%%%
where we discarded a divergent $a$-independent and $a_{0}$ is a non vanishing arbitrary constant, which does not contribute to the force.

	The case where $d=1$, $D=2$ and $N=2$ corresponds to the situation of two delta currents concentrated on parallel planes. In this situation the energy per unit of area (\ref{calEphicommassa1}) becomes
%%%%%%%%%
\begin{eqnarray}
{\cal E}_{I}(a,m,d=1,D=2,N=2)&=&-\frac{\sigma_{1}\sigma_{2}}{2m}\exp{(-ma)}\ ,
\end{eqnarray}
%%%%%%%%%
which reads for vanishing mass 
%%%%%%%%%
\begin{eqnarray}
\label{asd6}
{\cal E}_{I}(a,m=0,d=1,D=2)&=&\frac{\sigma_{1}\sigma_{2}}{2}a\ ,
\end{eqnarray}
%%%%%%%%%
after discarding a divergent $a$-independent term.

%%%%%%%%%
\section{Scalar Field in $d+D+1$ dimensions with delta-like derivative currents: dipole distributions}
\label{2}
%%%%%%%%%

	In this section we study the model described by the Lagrangian density
%%%%%%%%%
\begin{equation}
\label{modelo2}
{\cal L}_{II}=\frac{1}{2}(\partial_{\mu}\phi)(\partial^{\mu}\phi)-\frac{1}{2}m^{2}\phi^{2}+\Biggl(\partial^{\mu}\sum_{p=1}^{N}\sigma_{p}\ V_{\mu(p)}\ \delta^{d}({\bf x}_{\perp}-{\bf a}_{p})\Biggr)\phi\ ,
\end{equation}
%%%%%%%%%
where $V^{\mu}_{(p)}$ designates a fixed and static four vector in the reference frame we are performing the calculations.

	The Lagrangian (\ref{modelo2}) can be interpreted as the scalar field coupled to the current
%%%%%%%%%
\begin{equation}
\label{defJII}
J_{II}(x)=\partial^{\mu}\sum_{p=1}^{N}\sigma_{p}\ V_{\mu(p)}\ \delta^{d}({\bf x}_{\perp}-{\bf a}_{p})\ .
\end{equation}
%%%%%%%%%

	Following similar arguments which led to equation (\ref{asd1}) and performing two integrations by parts we have that the energy of the system (\ref{modelo2}) is given by
%%%%%%%%%
\begin{equation}
\label{asd7}
E_{II}=\sum_{p=1}^{N}\sum_{q=1}^{N}\sigma_{p}\sigma_{q}(1-\delta_{pq})\frac{1}{2T}{\cal J}_{pq}\ ,
\end{equation}
%%%%%%%%%
where ${\cal J}_{pq}$ is the integral defined by
%%%%%%%%%
\begin{equation}
\label{defcalJ}
{\cal J}_{pq}=\int\int d^{D+d+1}x\ d^{D+d+1}y\ \delta^{(d)}({\bf x}_{\perp}-{\bf a}_{p})\biggl(V^{\mu}_{(p)}V^{\nu}_{(q)}\partial_{\mu}\partial_{\nu}\ \Delta_{F}(x,y)\biggr)\delta^{(d)}({\bf y}_{\perp}-{\bf a}_{q})\ .
\end{equation}
%%%%%%%%%

	In the appendix A it is shown that
%%%%%%%%%
\begin{eqnarray}
\label{AresultadocalJ}
{\cal J}_{pq}=\Bigl({\bf V}_{(p)\perp}\cdot{\bf\nabla}_{pq\perp}\Bigr)\Bigl({\bf V}_{(q)\perp}\cdot{\bf\nabla}_{pq\perp}\Bigr){\cal I}_{pq}\ ,
\end{eqnarray}
%%%%%%%%%
where ${\bf V}_{(q)\perp}=(V_{(q)}^{1},...,V_{(q)}^{d})$ and
%%%%%%%%%
\begin{equation}
\label{nabla}
{\bf\nabla}_{pq\perp}=\left(\frac{\partial}{\partial a_{pq}^{1}},...,\frac{\partial}{\partial a_{pq}^{d}}\right)\ ,
\end{equation}
%%%%%%%%%
with $a_{pq}^{i}$ designating the $i$-th component of the vector ${\bf a}_{pq}$.

	By substituting Eq. (\ref{AresultadocalJ}) in (\ref{asd7}) and with the aid of (\ref{AresultadocalI}), (\ref{ext1}) and (\ref{defG}) we have finally
%%%%%%%%%
\begin{eqnarray}
\label{qwe2}
{\cal E}_{II}&=&\frac{1}{2}\sum_{p=1}^{N}\sum_{q=1}^{N}\sigma_{p}\sigma_{q}(1-\delta_{pq})\Bigl({\bf V}_{(p)\perp}\cdot{\bf\nabla}_{pq\perp}\Bigr)\Bigl({\bf V}_{(q)\perp}\cdot{\bf\nabla}_{pq\perp}\Bigr)\frac{1}{(2\pi)^{d/2}}m^{d-2}G(ma_{pq})\nonumber\\
&=&-\frac{1}{2}\frac{m^{d}}{(2\pi)^{d/2}}\sum_{p=1}^{N}\sum_{q=1}^{N}\sigma_{p}\sigma_{q}(1-\delta_{pq})\Biggl[(ma_{pq})^{-d/2}K_{d/2}(ma_{pq})\Bigl({\bf V}_{(p)\perp}\cdot{\bf V}_{(q)\perp}\Bigr)\nonumber\\
&\ &-(ma_{pq})^{-1-(d/2)}K_{1+(d/2)}(ma_{pq})\Bigl({\bf V}_{(p)\perp}\cdot(m{\bf a}_{pq})\Bigr)\Bigl({\bf V}_{(q)\perp}\cdot(m{\bf a}_{pq})\Bigr)\Biggr]\ ,
\end{eqnarray}
%%%%%%%%%

	For the massless case and $d\not=2$ we proceed in a similar way as we have done to obtain the energy (\ref{qwe2}) but, instead of (\ref{ext1}), we use the result (\ref{ext2}) what leads to
%%%%%%%%%
\begin{eqnarray}
\label{qwe3}
{\cal E}_{II}=\frac{2^{(d/2)-2}}{(2\pi)^{d/2}}\Gamma(d/2)\sum_{p=1}^{N}\sum_{q=1}^{N}\sigma_{p}\sigma_{q}(1-\delta_{pq})\frac{1}{a^{d}_{pq}}\Biggl[d\Biggl(\frac{{\bf V}_{(p)\perp}\cdot{\bf a}_{pq}}{a_{pq}}\Biggr)\Biggl(\frac{{\bf V}_{(q)\perp}\cdot{\bf a}_{pq}}{a_{pq}}\Biggr)-{\bf V}_{(p)\perp}\cdot{\bf V}_{(q)\perp}\Biggr]\ ,\  d\not=2\ .
\end{eqnarray}
%%%%%%%%%

	The situation where $m=0$ and $d=2$ is obtained by taking the limit $m\to0$ of (\ref{qwe2}) with $d=2$, what is done with the aid of (\ref{Km->0}), as follows
%%%%%%%%%
\begin{eqnarray}
\label{qwe3d=2}
{\cal E}_{II}
&=&-\frac{1}{4\pi}\sum_{p=1}^{N}\sum_{q=1}^{N}\sigma_{p}\sigma_{q}(1-\delta_{pq})\Bigl({\bf V}_{(p)\perp}\cdot{\bf\nabla}_{pq\perp}\Bigr)\Bigl({\bf V}_{(q)\perp}\cdot{\bf\nabla}_{pq\perp}\Bigr)\ln\Biggr(\frac{a_{pq}}{a_{0}}\Biggr)\cr\cr
&=&\frac{1}{4\pi}\sum_{p=1}^{N}\sum_{q=1}^{N}\sigma_{p}\sigma_{q}(1-\delta_{pq})\frac{1}{a^{2}_{pq}}\Biggl[2\Biggl(\frac{{\bf V}_{(p)\perp}\cdot{\bf a}_{pq}}{a_{pq}}\Biggr)\Biggl(\frac{{\bf V}_{(q)\perp}\cdot{\bf a}_{pq}}{a_{pq}}\Biggr)-{\bf V}_{(p)\perp}\cdot{\bf V}_{(q)\perp}\Biggr]\ ,\ d=2 \ ,
\end{eqnarray}
%%%%%%%%%
where, similarly to what was done in the previous section, in the first line of the above equation we discarded an $a_{pq}$-independent divergent term and introduced an arbitrary finite constant $a_{0}$ with length dimension.

	Notice that if we take $d=2$ in Eq. (\ref{qwe3}) we obtain (\ref{qwe3d=2}), it is, Eq. (\ref{qwe3}) can be considered to be valid with no restrictions for $d$.

	In order to have a better insight on the meaning of the current (\ref{defJII}), as well as of the energies (\ref{qwe2}) and (\ref{qwe3}), let us define ${\bf a}={\bf a}_{12}$, designate by ${\hat a}$ the unitary vector in the direction of ${\bf a}$ and consider the restricted situation for Eq. (\ref{qwe2}) where $N=2$ and $d=3$,
%%%%%%%%% 
\begin{eqnarray}
\label{E2commd=3}
{\cal E}_{II}(d=3,N=2)=\frac{\exp(-ma)}{4\pi a^{3}}\Biggl[\Bigl[(ma)^{2}+3(ma+1)\Bigr]\Bigl(\sigma_{1}{\bf V}_{\perp}^{(1)}\cdot{\hat a}\Bigr)\Bigl(\sigma_{2}{\bf V}_{\perp}^{(2)}\cdot{\hat a}\Bigr)-(ma+1)\Bigl(\sigma_{1}{\bf V}_{\perp}^{(1)}\cdot\sigma_{2}{\bf V}_{\perp}^{(2)}\Bigr)\Biggr]\ .
\end{eqnarray}
%%%%%%%%%
 An interesting situation occurs for the massless case, where equation (\ref{E2commd=3}) reads
%%%%%%%%%
\begin{equation}
\label{vfr}
{\cal E}=-\frac{1}{4\pi a^{3}}\Bigl[\Bigr(\sigma_{1}{\bf V}_{(1)}\cdot\sigma_{2}{\bf V}_{(2)}\Bigr)-3\Bigl(\sigma_{1}{\bf V}_{(1)}\cdot{\hat a}\Bigr)\Bigl(\sigma_{2}{\bf V}_{(2)}\cdot{\hat a}\Bigr)\Bigr]\ .
\end{equation}
%%%%%%%%%

	Notice that Eq. (\ref{vfr}) is exactly the interaction energy, with the negative sign, between two electric dipoles whose electric dipole moments are given by $\sigma_{1}{\bf V}_{(1)}$ and $\sigma_{2}{\bf V}_{(2)}$.

	The energy (\ref{qwe2}) is the generalization of the interaction between the dipoles distributions of the scalar field along $N$ $D$-dimensional parallel branes.
	
	It is worth mentioning that one could calculate the interaction between a dipole and a point-like charge by simulating them with currents of the form (\ref{defJII}) and (\ref{defJ}), respectively.

%%%%%%%%%
\section{Maxwell field}
\label{3,4}
%%%%%%%%%

	In this section we extend the models exposed in sections (\ref{1}) and (\ref{2}) to the Maxwell field, trying to be as general as possible.
	
	The first model we consider is described by the Maxwell field Lagrangian interacting with an external current, as follows
%%%%%%%%%
\begin{equation}
\label{qwe4}
{\cal L}_{III}=-\frac{1}{4}F_{\mu\nu}F^{\mu\nu}-\frac{1}{2\gamma}(\partial_{\mu}A^{\mu})^{2}+\biggl(\sum_{p=1}^{N}\sigma_{p}\  W_{\mu(p)}\ \delta^{d}({\bf x}_{\perp}-{\bf a}_{p})\biggr)A^{\mu}\ , 
\end{equation}
%%%%%%%%%
where $A^{\mu}$ is the photon field, $F^{\mu\nu}=\partial^{\mu}A^{\nu}-\partial^{\nu}A^{\mu}$ and $W^{\mu}_{(p)}$ is a four vector satisfying the condition ${\bf W}_{\perp}=0$ which assures that $\partial^{\mu}\Bigl[\sigma_{p}\ W_{\mu(p)}\ \delta^{d}({\bf x}_{\perp}-{\bf a}_{p})\Bigr]=0$ and so, the gauge invariance of the Lagrangian (\ref{qwe4}). The quantities $W^{\mu}_{(p)}$ shall be taken as constants and uniforms four vectors in the reference frame where the calculations are performed.

	Following similar steps of sections (\ref{1}) and (\ref{2}), and using the facts that ${\bf W}_{\perp}=0$  and the photon propagator is given by
%%%%%%%%%
\begin{equation}
\label{propfoton}
\Delta^{\mu\nu}(x,y)=-\int\frac{d^{D+d+1}k}{(2\pi)^{d+D+1}}\frac{1}{k^{2}}\Biggl[\eta^{\mu\nu}-(1-\gamma)\frac{k^{\mu}k^{\nu}}{k^{2}}\Biggr]\exp[-ik(x-y)]\ ,
\end{equation}
%%%%%%%%%
we can show that
%%%%%%%%%
\begin{equation}
\label{zxc1}
{\cal E}_{III}
=\frac{E_{3}}{L^{D}}=\frac{2^{(d/2)-1}}{(2\pi)^{d/2}}\Gamma\Biggl(\frac{d}{2}-1\Biggr)\sum_{p=1}^{N}\sum_{q=1}^{N}(1-\delta_{pq})\biggl(\sigma_{p}W_{(p)}^{\mu}\biggr)\biggl(\sigma_{q}W_{\mu(q)}\biggr)a_{pq}^{2-d}\ ,\ d\not=2\ .
\end{equation}
%%%%%%%%%

	For the case where $d=2$ we insert a mass parameter in the propagator (\ref{propfoton}) as a regulator parameter, as follows
%%%%%%%%%
\begin{equation}
\label{propfotoncommassa}
\Delta^{\mu\nu}(x,y;m)=-\int\frac{d^{D+d+1}k}{(2\pi)^{d+D+1}}\frac{1}{k^{2}-m^{2}}\Biggl[\eta^{\mu\nu}-(1-\gamma)\frac{k^{\mu}k^{\nu}}{k^{2}}\Biggr]\exp[-ik(x-y)]\ .
\end{equation}
%%%%%%%%%
It leads to the energy
%%%%%%%%%
\begin{eqnarray}
\label{zxc1d=2}
{\cal E}_{III}
&=&\frac{E_{III}}{L^{D}}=\frac{1}{4\pi}\sum_{p=1}^{N}\sum_{q=1}^{N}(1-\delta_{pq})\biggl(\sigma_{p}W_{(p)}^{\mu}\biggr)\biggl(\sigma_{q}W_{\mu(q)}\biggr)\lim_{m\to0}\Bigl(K_{0}(ma_{pq})\Bigr)\cr\cr
&=&-\frac{1}{4\pi}\sum_{p=1}^{N}\sum_{q=1}^{N}(1-\delta_{pq})\biggl(\sigma_{p}W_{(p)}^{\mu}\biggr)\biggl(\sigma_{q}W_{\mu(q)}\biggr)\ln\Biggl(\frac{a_{pq}}{a_{0}}\Biggr)\ ,\ d=2\ ,
\end{eqnarray}
%%%%%%%%%
where, in the second line, we used the second equation (\ref{Km->0}), discarded an $a_{pa}$-independent divergent term and $a_{0}$ is a finite arbitrary constant.

	By setting $N=2$, $d=3$, $D=0$ and $W^{\mu}=\eta^{\mu0}$ in Eq. (\ref{zxc1}) we have the Coulombian interaction between two charges in $3+1$ dimensions, whose energy is
%%%%%%%%%
\begin{equation}
E=\frac{\sigma_{1}\sigma_{2}}{4\pi}\frac{1}{a}\ ,
\end{equation}
%%%%%%%%%
with $a$ designating the distance between the charges.

	The second model we shall consider for the Maxwell field is an extension of the one considered in section (\ref{2}) for the scalar field. Its Lagrangian density is given by
%%%%%%%%%
\begin{equation}
\label{qwe5}
{\cal L}_{IV}=-\frac{1}{4}F_{\mu\nu}F^{\mu\nu}-\frac{1}{2\gamma}(\partial_{\mu}A^{\mu})^{2}+\sum_{p=1}^{N}\sigma_{p}\  W_{\mu(p)}\ V^{\alpha}_{(p)}\ \partial_{\alpha}\Bigl[\delta^{d}({\bf x}_{\perp}-{\bf a}_{p})\Bigr]\ .
\end{equation}
%%%%%%%%%

	Similarly to what happens for the model (\ref{qwe4}), the same condition ${\bf W}_{\perp}=0$ assures the gauge invariance of (\ref{qwe5}). The quantities $V^{\alpha}_{(p)}$ are four vector with no restrictions and, as well as $W_{\mu(p)}$, are taken to be static and uniform in the reference frame where the calculations are performed.

	By similar procedures which lead to results (\ref{qwe3}), (\ref{qwe3d=2}), (\ref{zxc1}) and (\ref{zxc1d=2}), the Lagrangian (\ref{qwe5}) leads to the energy per unit of parallel area $L^{D}$
%%%%%%%%%
\begin{eqnarray}
\label{qwe6}
{\cal E}_{IV}&=&\frac{2^{(d/2)-2}}{(2\pi)^{d/2}}\Gamma\Biggl(\frac{d}{2}\Biggr)\sum_{p=1}^{N}\sum_{q=1}^{N}(1-\delta_{pq})\Bigl(\sigma_{p}W_{(p)}^{\mu}\Bigr)\Bigl(\sigma_{q}W_{(q)\mu}\Bigr)\cr\cr
&\ &\times\frac{1}{a_{pq}^{d}}\Biggl[{\bf V}_{(p)\perp}\cdot{\bf V}_{(q)\perp}-d\Biggl(\frac{{\bf V}_{(p)}\cdot{\bf a}_{pq}}{a_{pq}}\Biggr)\Biggl(\frac{{\bf V}_{(q)}\cdot{\bf a}_{pq}}{a_{pq}}\Biggr)\Biggr]\ ,
\end{eqnarray}
%%%%%%%%%
which is valid for any value of $d$.

	The result (\ref{qwe6}) can be interpreted as the interaction energy between static electric dipoles distributions along $N$ $D$-dimensional parallel branes. This fact can be seen by considering the most interesting situation, which occurs for $D=0$, $d=3$, $N=2$ and $W^{\mu}_{(p)}=W^{\mu}_{(q)}=\eta^{\mu 0}$, where equation (\ref{qwe6}) reads
%%%%%%%%%
\begin{equation}
E_{IV}=\frac{1}{4\pi}\frac{1}{a^{3}}\Biggl[{(\sigma_{1}\bf V}_{(1)})\cdot(\sigma_{2}{\bf V}_{(2)})-3\Biggl(\frac{(\sigma_{1}{\bf V}_{(1)})\cdot{\bf a}}{a}\Biggr)\Biggl(\frac{(\sigma_{2}{\bf V}_{(2)})\cdot{\bf a}}{a}\Bigg)\Biggr]\ ,
\end{equation}
%%%%%%%%%
which is the interaction energy between two electric dipoles in $3+1$ dimensions separated by the vector ${\bf a}$ and with electric dipole moments given by $\sigma_{1}{\bf V}_{(1)}$ and $\sigma_{2}{\bf V}_{(2)}$.

	By these means each term in the summation present in (\ref{qwe5}) can be used to simulate the presence of a static electric dipole distribution along a brane with $D$ dimensions.

%%%%%%%%%
\section{Dirac field}
\label{5}
%%%%%%%%%

	In this section we consider the interaction of the Dirac's field, $\psi$, in $d+D+1$ dimensions with an external static current concentrated at some regions of the space. The Lagrangian density of the model is given by
%%%%%%%%%
\begin{equation}
\label{modelo5}
{\cal L}_{V}=\bar{\psi} (i\gamma^{\mu}\partial_{\mu}-m)\psi+\Biggl(\sum_{p=1}^{N}\sigma_{p}\ \bar{\xi}_{(p)}\delta^{d}({\bf x}_{\perp}-{\bf a}_{p})\Biggr)\psi+\bar{\psi}\Biggl(\sum_{q=1}^{N}\sigma_{q}\ \xi_{(q)}\delta^{d}({\bf x}_{\perp}-{\bf a}_{q})\Biggr)\ ,
\end{equation}
%%%%%%%%%
where $\xi_{(p)}$ are $N$ fixed and static spinors in our referential frame, $\gamma^{\mu}$ are the Dirac matrices and, as usual, $\bar{\xi}_{(p)}=\xi^{\dagger}_{(p)}\gamma^{0}$.

	Notice that the Lagrangian (\ref{modelo5}) can be interpreted as the Dirac's field coupled to the external fermionic current
%%%%%%%%%
\begin{equation}
\sum_{p=1}^{N}\sigma_{p}\ \xi_{(p)}\delta^{d}({\bf x}_{\perp}-{\bf a}_{p})\ .
\end{equation}
%%%%%%%%%

	Using the fact that the functional generator of the model (\ref{modelo1}) is given by
%%%%%%%%%
\begin{equation}
{\cal Z}=\exp{\Biggl[-i\int\int d^{d+D+1}x\ d^{d+D+1}y \Biggl(\sum_{p=1}^{N}\sigma_{p}\ {\bar\xi}_{(p)}^{*}\delta^{d}({\bf x}_{\perp}-{\bf a}_{p})\Biggr)S_{f}(x,y)\Biggl(\sum_{q=1}^{N}\sigma_{q}\ \xi_{(q)}\delta^{d}({\bf x}_{\perp}-{\bf a}_{q})\Biggr)\Biggr]}\ ,
\end{equation}
%%%%%%%%%
where
%%%%%%%%%
\begin{eqnarray}
S_{f}(x,y)&=&(i\gamma^{\mu}\partial_{\mu}+m)\int\frac{d^{d+D+1}k}{(2\pi)^{d+D+1}}\frac{1}{k^{2}-m^{2}}\exp{[-ik(x-y)]}\cr\cr
&=&\int\frac{d^{d+D+1}k}{(2\pi)^{d+D+1}}\frac{\gamma^{\mu}k_{\mu}+m}{k^{2}-m^{2}}\exp{[-ik(x-y)]}
\end{eqnarray}
%%%%%%%%%
is the usual Dirac propagator, and proceeding as in the previous cases, it can be shown that the field energy per unit of parallel area is given by
%%%%%%%%%
\begin{equation}
\label{poi1}
{\cal E}_{V}=\frac{E_{V}}{L^{D}}=\sum_{p=1}^{N}\sum_{q=1}^{N}\sigma_{p}^{*}\sigma_{q}(1-\delta_{pq})\ \bar{\xi}_{(p)}\left[\int\frac{d^{d}k}{(2\pi)^{d}}\frac{{\vec\gamma}\cdot{\bf k}_{\perp}-m}{{\bf k}_{\perp}^{2}+m^{2}}\exp{[i{\bf k}_{\perp}\cdot({\bf a}_{pq})]}\right]\xi_{(q)}
\end{equation}
%%%%%%%%%

	Defining the spatial vector
%%%%%%%%%
\begin{equation}
\label{defUdefvecgamma}
i{\bf U}_{pq}=\bar{\xi}_{(p)}{\vec\gamma}\ \xi_{(q)}\ \ \ \ {\vec\gamma}=(\gamma^{1},\cdot\cdot\cdot,\gamma^{d+D})\ ,
\end{equation}
%%%%%%%%%
equation (\ref{poi1}) becomes
%%%%%%%%%
\begin{eqnarray}
\label{poi2}
{\cal E}_{V}&=&\sum_{p=1}^{N}\sum_{q=1}^{N}\sigma_{p}^{*}\sigma_{q}(1-\delta_{pq})\Bigl[{\bf U}_{pq}\cdot{\bf\nabla}_{pq\perp}-m{\bar\xi}_{(p)}\xi_{(q)}\Bigr]\int\frac{d^{d}k}{(2\pi)^{d}}\frac{1}{{\bf k}_{\perp}^{2}+m^{2}}\exp{[i{\bf k}_{\perp}\cdot({\bf a}_{pq})]}
\cr\cr
&=&(2\pi)^{-d/2}m^{d-2}\sum_{p=1}^{N}\sum_{q=1}^{N}\sigma_{p}^{*}\sigma_{q}(1-\delta_{pq})\Bigl[{\bf U}_{pq}\cdot{\bf\nabla}_{pq\perp}-m{\bar\xi}_{(p)}\xi_{(q)}\Bigr]G(ma_{pq})\ ,
\end{eqnarray}
%%%%%%%%%
where, in the second line, we used the result (\ref{ext1}) and definition (\ref{defG}).

	With the aid of (\ref{defG}) equation (\ref{poi2}) reads
%%%%%%%%%
\begin{eqnarray}
\label{poi3}
{\cal E}_{V}&=&-(2\pi)^{-d/2}m^{d-1}\sum_{p=1}^{N}\sum_{q=1}^{N}(1-\delta_{pq})(ma_{pq})^{1-(d/2)}\Biggl[K_{d/2}(ma_{pq})\Biggl((\sigma_{p}^{*}{\bf U}_{pq}\sigma_{q})\cdot\frac{{\bf a}_{pq}}{a_{pq}}\Biggr)\cr\cr
&\ &+K_{(d/2)-1}(ma_{pq})\Bigr[(\sigma_{p}^{*}{\bar\xi}_{(p)})(\sigma_{q}\xi_{(q)})\Bigr]\Biggr]\ .
\end{eqnarray}
%%%%%%%%%

	Taking into account that ${\bf a}_{pq}=-{\bf a}_{qp}$\ , $a_{pq}=a_{qp}$, $(\sigma_{p}^{*}{\bar\xi}_{(p)})(\sigma_{q}\xi_{(q)})=\Bigl[(\sigma_{q}^{*}{\bar\xi}_{(q)})(\sigma_{p}\xi_{(p)})\Bigr]^{*}$ and 
$(\sigma_{p}^{*}{\bf U}_{pq}\sigma_{q})=-(\sigma_{q}^{*}{\bf U}_{qp}\sigma_{p})^{*}$, equation (\ref{poi3}) can be written  in the form
%%%%%%%%%
\begin{eqnarray}
\label{energia5}
{\cal E}_{V}&=&-\frac{m^{d-1}}{(2\pi)^{d/2}}\sum_{p=1}^{N}\sum_{q=1}^{N}(1-\delta_{pq})(ma_{pq})^{1-(d/2)}\Biggl[K_{d/2}(ma_{pq})\ \Im\Bigl[(\sigma_{p}^{*}\bar{\xi}_{(p)}){\vec\gamma}\ (\sigma_{q}\xi_{(q)})\Bigr]\cdot\frac{{\bf a}_{pq}}{a_{pq}}\cr\cr
&\ &+K_{(d/2)-1}(ma_{pq})\ \Re\Bigl[(\sigma_{p}^{*}{\bar\xi}_{(p)})(\sigma_{q}\xi_{(q)})\Bigr]\Biggr]\ ,
\end{eqnarray}
%%%%%%%%%
where we used definition (\ref{defUdefvecgamma}).

	For the massless case with $d=3$, $D=0$ and $N=2$ the energy (\ref{energia5}) reads
%%%%%%%%%
\begin{equation}
{\cal E}_{5}(m=0,d=3)=E_{5}(m=0,d=3)=-\frac{1}{2\pi}\frac{1}{a^{2}}\frac{{\bf a}}{a}\cdot\Im\Bigl[(\sigma_{1}^{*}\bar{\xi}_{(1)}){\vec\gamma}\ (\sigma_{2}\xi_{(2)})\Bigr]\ ,
\end{equation}
%%%%%%%%%
where $a=a_{12}$, which leads to the force acting on the current 2
%%%%%%%%%
\begin{equation}
\label{forcafermion}
{\bf F}=\frac{1}{2\pi}\frac{1}{a^{3}}\Biggl[3{\hat a}\Bigl[{\hat a}\cdot\Im\Bigl((\sigma_{1}^{*}\bar{\xi}_{(1)}){\vec\gamma}\ (\sigma_{2}\xi_{(2)})\Bigr)\Bigr]-\Im\Bigl((\sigma_{1}^{*}\bar{\xi}_{(1)}){\vec\gamma}\ (\sigma_{2}\xi_{(2)})\Bigr)\Biggr]\ .
\end{equation}
%%%%%%%%%

	It is interesting to notice that the force field (\ref{forcafermion}) has the same spacial behavior as the one exhibited by the electric field produced by a static electric dipole given by $\Im\Bigl((\sigma_{1}^{*}\bar{\xi}_{(1)}){\vec\gamma}\ (\sigma_{2}\xi_{(2)})\Bigr)$.

%%%%%%%%%
\section{$\lambda\phi^4$ model with two delta-like currents: confining potentials}
\label{lambdaphi4}
%%%%%%%%%

	In this section we consider the radiative corrections induced by the, well known, $\lambda\phi^4$ self-interaction on the energy of the scalar field interacting with static external currents in $3+1$ dimensions. 
		
	The Lagrangian of the model is written, in the general form,
%%%%%%%%%
\begin{equation}
{\cal L}=\frac{1}{2}(\partial_{\mu}\phi)(\partial^{\mu}\phi)-\frac{1}{2}m^{2}\phi^{2}-\frac{\lambda}{4!}\phi^4-\frac{1}{2}\delta m^{2}\phi^{2}+J\phi\ ,
\end{equation}
%%%%%%%%%
where $\delta m^{2}$ is a mass counter term 
and the current depends only on the perpendicular coordinates ${\bf x}_{\perp}$.
Once we are restricted to $3+1$ dimensions, we must always have $d+D=3$.

	By using standard perturbative methods of quantum field theory we write the functional generator ${\cal Z}(J)$ up to first order in the coupling constant $\lambda$
%%%%%%%%%
\begin{equation}
\label{fgenerator}
{\cal Z}(J)=\exp\left[-i\int d^4 x \left(-\frac{\delta m^{2}}{2}\frac{\delta^{2}}{\delta J(x)^{2}}\right)+\left(\frac{\lambda}{4!}\frac{\delta^{4}}{\delta J(x)^{4}}\right)\right]
\exp\Biggl(-\frac{i}{2}\int\ d^4x\ d^4y\ \ J(x)\Delta_{F}(x,y)J(y)\Biggr)\ .
\end{equation}
%%%%%%%%%
From now on we shall consider the restricted case where the current $J$ is located at two distinct regions of the space.

	Substituting the functional (\ref{fgenerator}) in Eq. (\ref{FG}) we have, at first order in $\lambda$,
%%%%%%%%%
\begin{eqnarray}
\label{firstorder}
E&=&\lim_{T\to\infty}\frac{1}{2T}\int d^{d+D+1}x\ d^{d+D+1}y\ J(x)\Delta_{F}(x-y)J(y)\cr\cr
&\ &+\lim_{T\to\infty}\frac{1}{2T}\left(\frac{\delta m^{2}}{2}+\frac{6i\lambda}{4!}\Delta_{F}(0)\right)\times\cr\cr
&\ &\int d^{d+D+1}x\ d^{d+D+1}y\ d^{d+D+1}z\ 
J(x)\Delta_{F}(x-y)\Delta_{F}(y-z)J(z)
\end{eqnarray}
%%%%%%%%%
where we disregarded non quadratic terms in $J$ in order to avoid contributions to the energy due to self interactions, as shall be exposed later.

	The first term on the right hand side of Eq. (\ref{firstorder}) gives the contribution to the energy in zero order in $\lambda$ studied in sections (\ref{1}) and (\ref{2}). The second term gives the correction in order $\lambda$ to the energy and, by the use of (\ref{defDeltaF}) and after the integrations $d^{d+D+1}x\ d^{D+1}y_{\|}\ d^{D+1}z_{\|} d^{D+1}k_{\|}$, can be simplified to
%%%%%%%%%
\begin{eqnarray}
\label{abgt1}
\frac{E^{(1)}}{L^{D}}=\frac{1}{2}\left(\frac{\delta m^{2}}{2}+\frac{6i\lambda}{4!}\Delta_{F}(0)\right)\int\ d^{d}{\bf y}_{\perp}\ d^{d}{\bf z}_{\perp} J({\bf y}_{\perp})J({\bf z}_{\perp})\int\frac{d^{d}{\bf k}_{\perp}}{(2\pi)^{d}}\frac{\exp[-i{\bf k}_{\perp}\cdot({\bf y}_{\perp}-{\bf z}_{\perp})]}{({\bf k}_{\perp}^{2}+m^{2})^{2}}\ .
\end{eqnarray}
%%%%%%%%%
In above expression, the counter-term $\delta m^{2}$ shall be defined for each specific situation considered later.
	
	Now, let us take an external current with in the form (\ref{defJ}), but with only two Dirac's delta functions present,
%%%%%%%%% 
\begin{equation}
\label{abgt2}
J({\bf x}_{\perp})=\sigma_{1}\delta^{d}({\bf x}_{\perp}-{\bf a}_{1})+\sigma_{2}\delta^{d}({\bf x}_{\perp}-{\bf a}_{2})\ .
\end{equation}
%%%%%%%%%

	Substituting (\ref{abgt2}) in (\ref{abgt1}), performing the integration in $d^{d}{\bf y}_{\perp}\ d^{d}{\bf z}_{\perp}$ and discarding the self interacting terms between a given delta with itself we have
%%%%%%%%%
\begin{eqnarray}
\label{abgt3}
{\cal E}^{(1)}&=&\frac{E^{(1)}}{L^{D}}=\left(\frac{\delta m^{2}}{2}+\frac{6i\lambda}{4!}\Delta_{F}(0)\right)\sigma_{1}\sigma_{2}\int\frac{d^{d}{\bf k}_{\perp}}{(2\pi)^{d}}\frac{\exp(-i{\bf k}_{\perp}\cdot{\bf a})}{({\bf k}_{\perp}^{2}+m^{2})^{2}}\cr\cr
&=&-\frac{1}{2}\left(\frac{\delta m^{2}}{2}+\frac{6i\lambda}{4!}\Delta_{F}(0)\right)\frac{\sigma_{1}\sigma_{2}}{m}\frac{\partial}{\partial m}\int\frac{d^{d}{\bf k}_{\perp}}{(2\pi)^{d}}\frac{\exp(-i{\bf k}_{\perp}\cdot{\bf a})}{{\bf k}_{\perp}^{2}+m^{2}}\ ,
\end{eqnarray}
%%%%%%%%%%
where we defined ${\bf a}={\bf a}_{1}-{\bf a}_{2}$.

	Contributions for the energy (\ref{firstorder}) quadratic in $J$ give rise only to self interacting terms for the deltas, and must be discarded.   
	Using the result (\ref{ext1}) Eq. (\ref{abgt3}) gives, for the case with mass,
%%%%%%%%%
\begin{eqnarray}
\label{abgt4}
{\cal E}^{(1)}(m,d,D,a)&=&-\frac{1}{2}\left(\frac{\delta m^{2}}{2}+\frac{6i\lambda}{4!}\Delta_{F}(0)\right)\frac{\sigma_{1}\sigma_{2}}{(2\pi)^{d/2}}a^{4-d}(ma)^{(d/2)-2}\times\cr\cr
&\ &\Bigl[-K_{d/2}(ma)+(d-2)(ma)^{-1}K_{(d/2)-1}(ma)\Bigr]
\end{eqnarray}
%%%%%%%%%

	The counter-term $\delta m^{2}$ is determinated by imposing that the above expression is finite, what is equivalent of taking
%%%%%%%%% 
\begin{equation}
\label{contratermo1}
\frac{1}{2}\left(\frac{\delta m^{2}}{2}+\frac{6i\lambda}{4!}\Delta_{F}(0)\right)=\lambda\kappa(m,d)
\end{equation}
%%%%%%%%%
where $\kappa(m,d)$ is a given finite parameter which must be determinated experimentally, can depend on $m$ and $d$ and whose dimension is mass square. So, Eq. (\ref{abgt4}) reads 
%%%%%%%%%
\begin{eqnarray}
{\cal E}^{(1)}(m,d,D,a)&=&-\lambda\kappa(m,d)\frac{\sigma_{1}\sigma_{2}}{(2\pi)^{d/2}}a^{4-d}(ma)^{(d/2)-2}\times\cr\cr
&\ &\Bigl[-K_{d/2}(ma)+(d-2)(ma)^{-1}K_{(d/2)-1}(ma)\Bigr].
\end{eqnarray}
%%%%%%%%%
	
	For $d=3$, $d=2$ and $d=1$ we have, respectively
%%%%%%%%%
\begin{eqnarray}
{\cal E}^{(1)}(m,d=3,D=0,a)&=&\lambda\kappa(m,d=3)\frac{\sigma_{1}\sigma_{2}}{4\pi}\frac{\exp{(-ma)}}{m}\cr\cr
{\cal E}^{(1)}(m,d=2,D=1,a)&=&\lambda\kappa(m,d=2)\frac{\sigma_{1}\sigma_{2}}{2\pi}a^{2}\frac{K_{1}{(ma)}}{ma}\cr\cr
{\cal E}^{(1)}(m,d=1,D=2,a)&=&\lambda\kappa(m,d=1)\frac{\sigma_{1}\sigma_{2}}{2}a^{3}(1+ma)\frac{1}{(ma)^3}\exp(-ma)
\end{eqnarray}
%%%%%%%%%

	The situation where $m=0$ can be obtained by taking the limit of vanishing mass in above results, what can be easily done for $d\not=2$. In this paper we avoid  the case $m=0$ and $d=2$, where infrared divergences are problematic. 
	
For $d\not=2$ we take $m=0$ in the first line of Eq. (\ref{abgt3}) and use the fact that
%%%%%%%%%
\begin{eqnarray}
\label{Ilambda}
\int{d^{d}{\bf p}}\frac{1}{p^{4}}\exp(\pm i{\bf p}{\bf a})&=&(2\pi)^{d/2}a^{4-d}\int_{0}^{\infty}du\ u^{(d/2)-4}J_{(d/2)-1}(u)\cr\cr
&=&(2\pi)^{d/2}a^{4-d}2^{(d/2)-4}\Gamma\Biggl(\frac{d}{2}-2\Biggr)\ ,\ d\not=2\ ,
\end{eqnarray}
%%%%%%%%%
which is obtained in appendix B. So we have
%%%%%%%%%
\begin{equation}
\label{abgt5}
{\cal E}^{1}=\left(\frac{\delta m^{2}}{2}+\frac{6i\lambda}{4!}\Delta_{F}(0)\right)\frac{\sigma_{1}\sigma_{2}}{(2\pi)^{d/2}}a^{4-d}2^{(d/2)-2}\Gamma\Biggl(\frac{d}{2}-2\Biggr)\ ,\ d\not=2\ .
\end{equation}
%%%%%%%%%

	For $d=3$ and $d=1$ the counter-term $\delta m^{2}$ is determinate by (\ref{contratermo1}) and we have
%%%%%%%%%
\begin{eqnarray}
\label{abgt6}
{\cal E}^{(1)}(m=0,d=3,D=0,a)&=&-\lambda\kappa(m=0,d=3)\frac{\sigma_{1}\sigma_{2}}{4\pi}a\cr\cr
{\cal E}^{(1)}(m=0,d=1,D=2,a)&=&\lambda\kappa(m=0,d=1)\frac{\sigma_{1}\sigma_{2}}{6}a^{3}\ .
\end{eqnarray}
%%%%%%%%%	
	
	We would like to point out that, depending on the signals of the parameters $\kappa$ and $\kappa'$, the expressions given by (\ref{abgt6}) can be confining potentials.
		
	Another interesting current is the one considered in section (\ref{2}) which we shall take, in this section, to be composed only by  two terms, that is, we shall take $N=2$ in (\ref{modelo2}), 
%%%%%%%%%
\begin{equation}
\label{edc1}
J({\bf x}_{\perp})=\sigma_{1}V^{\mu}_{(1)}\partial_{\mu}[\delta({\bf x}_{\perp}-{\bf a}_{1})]+\sigma_{2}V^{\mu}_{(2)}\partial_{\mu}[\delta({\bf x}_{\perp}-{\bf a}_{2})]\ .
\end{equation}
%%%%%%%%%

	Substituting (\ref{edc1}) in (\ref{abgt1}) we have, after some manipulations similar to the ones done in the previous sections,
%%%%%%%%%
\begin{equation}
\label{edc2}
{\cal E}^{(1)}=-\left(\frac{\delta m^{2}}{2}+\frac{6i\lambda}{4!}\Delta_{F}(0)\right)\sigma_{1}\sigma_{2}({\bf V}_{(1)\perp}\cdot{\bf \nabla}_{a})({\bf V}_{(2)\perp}\cdot{\bf \nabla}_{a})\int_{0}^{\infty}\frac{d^{d}{\bf k}_{\perp}}{(2\pi)^{d}}\frac{\exp(-i{\bf k}_{\perp}\cdot{\bf a})}{({\bf k}_{\perp}^{2}+m^{2})^{2}}\ .
\end{equation}
%%%%%%%%%	
The above integral is calculated in (\ref{abgt3}) and (\ref{abgt4}).

	The general result for (\ref{edc2}) is not much suggestive.	The most interesting situation occurs in the massless case for $d=3$, where we must take the counter term $\delta m^{2}$ in such a way that
%%%%%%%%%
\begin{equation}
\frac{1}{2}\left(\frac{\delta m^{2}}{2}+\frac{6i\lambda}{4!}\Delta_{F}(0)\right)=\lambda{\tilde\kappa}\ ,
\end{equation}
%%%%%%%%%
where $\tilde\kappa$ is a finite constant which must be determinated experimentally. So Eq. (\ref{edc2}) leads to
%%%%%%%%%
\begin{equation}
{\cal E}^{(1)}=\lambda{\tilde\kappa}\frac{\sigma_{1}\sigma_{2}}{4\pi}\frac{1}{a}\Bigl[{\bf V}_{(1)\perp}\cdot{\bf V}_{(2)\perp}-({\bf V}_{(1)\perp}\cdot{\hat a})({\bf V}_{(2)\perp}\cdot{\hat a})\Bigr]
\end{equation}
%%%%%%%%%
which is an anisotropic long range interaction energy. In fact, its angular dependence is the same one exhibited by the interaction between two static dipoles.

%%%%%%%%%
\section{Conclusions and Final Remarks}
\label{conclusao}
%%%%%%%%%

	In this paper we have made an investigation on the role of the interaction of quantum fields with external static currents concentrated along different parallel branes. Specifically we generalized to $d+D+1$ dimensions the interaction between bosonic fields (scalar and electromagnetic) with external currents composed by $N$ $d$-dimensional Dirac delta functions and delta functions derivatives. For the scalar case we considered the situations with and without mass. We showed that we can describe charges distributions as well as static dipole distributions by the use of currents concentrated along parallel branes with arbitrary codimensions.

	We also had made the same study for the fermionic field and, in the massless case, we showed that we have a dipole-like interaction.
		
	We showed that the first order radiative corrections induced by the $\lambda\phi^{4}$ self interaction, for the case of scalar field, can lead to confining potentials between point-like currents, as well as a long range anisotropic interaction.
	
	As a last comment we would like to say that four-pole interactions can also be described by the use of currents concentrated along specific regions of space \cite{proximo}.

\ 

\

{\bf Acknowledgements }

	The authors would like to thank C. Farina, J.A. Helay\"el-Neto and N.F Svaiter for discussions and suggestions and FAPEMIG for invaluable financial support.

%%%%%%%%%
\section*{Appendix A}
%%%%%%%%%

 In order to write the integral (\ref{AdefcalI}) in the form (\ref{AresultadocalI}) we, first, use definition (\ref{defDeltaF}), what leads to
%%%%%%%%%
\begin{eqnarray}
{\cal I}_{p,q}&=&\int\int d^{0}x\ d^{0}y\ \int\int d^{D}x_{\|}d^{D}y_{\|}\ \int\int d^{d}x_{\perp}d^{d}y_{\perp}\ \delta^{(d)}({\bf x}_{\perp}-{\bf a}_{p})\ \delta^{(d)}({\bf y}_{\perp}-{\bf a}_{q})\cr\cr
&\times& \int \frac{dk^{0}}{2\pi}\int\frac{d{\bf k}_{\perp}}{(2\pi)^{D}}\int\frac{d{\bf k}_{\|}}{(2\pi)^{d}}
\exp{[ik^{0}(x^{0}-y^{0})]}\exp{[-i{\bf k}_{\|}\cdot({\bf x}_{\|}-{\bf y}_{\|})]}\cr\cr
&\times&\frac{\exp{[-i{\bf k}_{\perp}\cdot({\bf x}_{\perp}-{\bf y}_{\perp})]}}{(k^{0})^{2}-{\bf k}_{\|}^{2}-{\bf k}_{\perp}^{2}-m^{2}}\ .\nonumber\\ 
\end{eqnarray}
%%%%%%%%%
Performing, in the following order, the integrals $d^{d}x_{\perp}$, $d^{d}y_{\perp}$, $d^{0}x$ and $d^{D}x_{\|}$, using the fact that
%%%%%%%%%
\begin{equation}
\delta(k)=\int\frac{dx}{2\pi}\exp(ikx)\ ,
\end{equation}
%%%%%%%%%
and integrating in the variables $dk^{0}$ and $d{\bf k}_{\|}$, we have
%%%%%%%%%
\begin{equation}
\label{Aasd1}
{\cal I}_{p,q}=-\int dy^{0}\ \int d^{D}{\bf y}_{\|}\ \int\frac{d^{d}{\bf k}_{\perp}}{(2\pi)^{d}}\ \frac{1}{{\bf k}_{\perp}^{2}+m^{2}}\ \exp{[-i{\bf k}_{\perp}\cdot({\bf a}_{p}-{\bf a}_{q})]}
\end{equation}
%%%%%%%%%

	As usually done in quantum field theory, we designate the integral in the time variable by $T=\int dy^{0}$. We also designate the integral in the parallel volume by
%%%%%%%%%
\begin{equation}
L^{D}=\int d^{D}{\bf y}_{\|}\ .
\end{equation}
%%%%%%%%%
With these considerations, Eq. (\ref{Aasd1}) is written in the form (\ref{AresultadocalI}).

	Another useful integral is the one defined in (\ref{defcalJ}) which, following similar steps used to obtain (\ref{AresultadocalI}), can be rewritten into the form
%%%%%%%%%
\begin{equation}
\label{bgt}
{\cal J}_{pq}=TL^{D}\int\frac{d^{d}{\bf k}_{\perp}}{(2\pi)^{d}}\left[\frac{\Bigl({\bf V}_{(p)\perp}\cdot{\bf k}_{\perp}\Bigr)\Bigl({\bf V}_{(q)\perp}\cdot{\bf k}_{\perp}\Bigr)}{k_{\perp}^{2}+m^{2}}\right]\exp(-i{\bf k}_{\perp}\cdot{\bf a}_{pq})
\end{equation}
%%%%%%%%%
where we defined ${\bf V}_{(q)\perp}=(V_{(q)}^{1},V_{(q)}^{2},...,V_{(q)}^{d})$.

	By using the operator (\ref{nabla}) Eq. (\ref{bgt}) becomes
%%%%%%%%%
\begin{eqnarray}
\label{bgt2}
{\cal J}_{pq}&=&-TL^{D}\Bigl({\bf V}_{(p)\perp}\cdot{\bf\nabla}_{pq\perp}\Bigr)\Bigl({\bf V}_{(q)\perp}\cdot{\bf\nabla}_{pq\perp}\Bigr)\int\frac{d^{d}{\bf k}_{\perp}}{(2\pi)^{d}}\frac{1}{k_{\perp}^{2}+m^{2}}\exp(-i{\bf k}_{\perp}\cdot{\bf a}_{pq})
\end{eqnarray}
%%%%%%%%%

	With the aid of (\ref{AresultadocalI}) Eq. (\ref{bgt2}) leads to (\ref{AresultadocalJ}).

%%%%%%%%%
\section*{Appendix B}
%%%%%%%%%

	In this appendix we calculate the $d$-dimensional integral
%%%%%%%%%
\begin{equation}
\label{defId}
I_{d}(a)=\int d^{d}{\bf p}\ f(p) \exp(\pm i{\bf p}{\bf a})\ ,
\end{equation}
%%%%%%%%%
where $f(p)$ is any function of the modulus $p$ of the vector ${\bf p}$ which satisfies the property $f(p)=f(-p)$.

	For this task let us consider, first, the case where $d>2$.
	
	It is appropriate to use spherical coordinates in $d$ dimensions, in such a way that
%%%%%%%%%
\begin{eqnarray}
p^{1}  &=&p\ \cos(\theta_{1})\nonumber\\
p^{2}  &=&p\ \sin(\theta_{1}) \cos(\theta_{2})\nonumber\\
p^{3}  &=&p\ \sin(\theta_{1}) \sin(\theta_{2}) \cos(\theta_{3})\nonumber\\
     &\vdots&\nonumber\\
p^{d-1}&=&p\ \sin(\theta_{1}) \sin(\theta_{2}) \cdots \sin(\theta_{d-2}) \cos(\theta_{d-1})\nonumber\\
p^{d}  &=&p\ \sin(\theta_{1}) \sin(\theta_{2}) \cdots \sin(\theta_{d-2}) \sin(\theta_{d-1})\ ,\nonumber
\end{eqnarray}
%%%%%%%%%
where $0\leq\theta_{1}\leq\pi$, $0\leq\theta_{2}\leq\pi$, ..., $0\leq\theta_{d-2}\leq\pi$ and $0\leq\theta_{d-1}\leq 2 \pi$. 
In this case we have
%%%%%%%%%
\begin{equation}
d^{d}{\bf p}=(p)^{d-1}\ dp\ \prod_{i=1}^{d-1}\sin^{d-(i+1)}(\theta_{i})\ d\theta_{i}\ .
\end{equation}
%%%%%%%%%

	It is also convenient to use a coordinate system in such a way that ${\bf a}=(a,0,\cdots,0)$. So, the integral (\ref{defId}) becomes
%%%%%%%%%
\begin{eqnarray}
\label{aqwe1}
I_{d}(a)&=&\Biggl[\int_{0}^{\infty}dp\ (p)^{d-1}f(p)\Biggl(\int_{0}^{\pi}d\theta_{1}\ \sin^{d-2}(\theta_{1})\ \exp\Bigl[\pm ipa\cos(\theta_{1})\Bigr]\Biggr)\Biggr]\nonumber\\
&\times&\Biggl(\int_{0}^{\pi}d\theta_{2}\ \sin^{d-3}(\theta_{2})\Biggr)\Biggl(\int_{0}^{\pi}d\theta_{3}\ \sin^{d-4}(\theta_{3})\Biggr)\cdots\Biggl(\int_{0}^{\pi}d\theta_{d-2}\ \sin^{d-3}(\theta_{d-2})\Biggr)\Biggl(\int_{0}^{2\pi}d\theta_{d-1}\Biggr)\ .\nonumber\\
\end{eqnarray}
%%%%%%%%%

	By using the results \cite{Arfken}
%%%%%%%%%
\begin{eqnarray}
\int_{0}^{\pi}\ d\theta \sin^{2\nu}(\theta)\ \exp\Bigl[\pm ix\cos(\theta)\Bigr]&=&\pi^{1/2}\Gamma\Biggl(\nu+\frac{1}{2}\Biggr)\Biggl(\frac{x}{2}\Biggr)^{-\nu}J_{\nu}(x)\ \ ,\ \ \nu>-\frac{1}{2}\nonumber\\ \nonumber\\
\int_{0}^{\pi}d\theta\ \sin^{m}(\theta)&=&\pi^{1/2}\frac{\Gamma[(1/2)(m+1)]}{\Gamma[(1/2)(m+2)]}\ \ \ \ \ ,\ \ \ \ m=1,2,3,...
\end{eqnarray}
%%%%%%%%%
equation (\ref{aqwe1}) reads, after some manipulations,
%%%%%%%%%
\begin{equation}
\label{Id>2}
I_{d}=(2\pi)^{d/2}\frac{1}{a^{d}}\int_{0}^{\infty}du\ u^{d/2} J_{(d/2)-1}(u)f\Biggl(\frac{u}{a}\Biggr)\ \ \ ,\ \ \ d>2\ ,
\end{equation}
%%%%%%%%%
where we made the change of integration variable $u=ap$.

	For $d=2$ dimensions, usual spherical coordinates leads to
%%%%%%%%%
\begin{equation}
I_{2}=\int_{0}^{\infty} dp\ p\ f(p)\ \int_{0}^{2\pi} d\theta\ \exp\Bigl[\pm ipa\cos(\theta_{1})\Bigr]\ .
\end{equation}
%%%%%%%%%
Noticing that
%%%%%%%%%
\begin{equation}
\int_{0}^{2\pi} d\theta \exp\Bigl[\pm ix\cos(\theta_{1})\Bigr]=2\pi J_{0}(x)
\end{equation}
%%%%%%%%%
and performing the change of integration variable $u=pa$ we have
%%%%%%%%%
\begin{equation}
\label{I2}
I_{2}=\frac{2\pi}{a^{2}}\int_{0}^{\infty}du\ u J_{0}(u) f\Biggl(\frac{u}{a}\Bigg)\ .
\end{equation}
%%%%%%%%%

	For $d=1$ dimensions we use the fact that $f(p)=f(-p)$ in order to write
%%%%%%%%%
\begin{eqnarray}
I_{1}&=&\int_{-\infty}^{\infty}dp\ f(p)\exp(\pm ipa)\nonumber\\
&=&\frac{2}{a}\int_{0}^{\infty}du\ f\Biggl(\frac{u}{a}\Biggr)\cos(u)\ ,
\end{eqnarray}
%%%%%%%%%
where we also made the change $u=pa$.

	Using that
%%%%%%%%%
\begin{equation}
\cos(u)=\sqrt{\frac{\pi}{2}}\sqrt{u}J_{-1/2}(u)\ , 
\end{equation}
%%%%%%%%%
we have
%%%%%%%%%
\begin{equation}
\label{I1}
I_{1}=\frac{(2\pi)^{1/2}}{a}\int_{0}^{\infty}du\ u^{1/2}J_{-1/2}(u)f\Biggl(\frac{u}{a}\Biggr)
\end{equation}
%%%%%%%%%

	Combining the results (\ref{Id>2}), (\ref{I2}) and (\ref{I1}) we are taken to the general result
%%%%%%%%%
\begin{eqnarray}
\label{Id}
I_{d}&=&\int d^{d}{\bf p}\ f(p) \exp(\pm i{\bf p}{\bf a})\nonumber\\
&=&(2\pi)^{d/2}\frac{1}{a^{d}}\int_{0}^{\infty}du\ u^{d/2} J_{(d/2)-1}(u)f\Biggl(\frac{u}{a}\Biggr)
\end{eqnarray}
%%%%%%%%%

	Applying formula (\ref{Id}) for the case where $f(p)=(p^2+m^2)^{-1}$ and using the fact that \cite{Gradstein}
%%%%%%%%%
\begin{equation}
\label{Bprincipal}
\int_{0}^{\infty}du\ \frac{u^{\nu+1}J_{\nu}(u)}{u^{2}+x^{2}}=x^{\nu}K_{\nu}(x)\ ,
\end{equation}
%%%%%%%%%
which is valid for $x>0$ and $-1<\nu<3/2$, we are taken to the result (\ref{Icommassa}).

	Taking $f(p)=p^{-2}$ in Eq. (\ref{Id}) and using the fact that \cite{Gradstein}
%%%%%%%%%
\begin{equation}
\label{Bsegunda}
\int_{0}^{\infty}du\ u^{\mu}J_{\nu}(u)=2^{\mu}\ \frac{\Gamma\bigl(1/2+\nu/2+\mu/2\bigr)}{\Gamma\bigl(1/2+\nu/2-\mu/2\bigr)}
\end{equation}
%%%%%%%%%
which is valid for $-\nu-1<\mu<1/2$, we obtain Eq. (\ref{Isemmassa}).

	Another case of interest corresponds to $f(p)=p^{-4}$ which, with the aid of (\ref{Bsegunda}), leads to (\ref{Ilambda}).

%%%%%%%%%

%%%%%%%%%

\end{document}